\documentclass[aps,prl,twocolumn,showpacs,superscriptaddress,groupedaddress]{revtex4}  % for review and submission
\usepackage{graphicx}  % needed for figures
\usepackage{dcolumn}   % needed for some tables
\usepackage{bm}        % for math
\usepackage{amssymb}   % for math
\usepackage{color}
\usepackage{hyperref}
\usepackage{epstopdf}
\usepackage{pdfpages}
\usepackage{soul}
\usepackage{natbib}
\hypersetup{
	colorlinks=true,
	urlcolor= blue,
	citecolor=blue,
	linkcolor= blue,
	bookmarks=true,
	bookmarksopen=false,
}

\begin{document}
\title{Maxwell's demon-like nonreciprocity by non-Hermitian gyrotropic metasurfaces} %Title of paper

\author{Wenyan Wang}
\affiliation{College of Physics and Optoelectronics, Key Lab of Advanced Transducers and Intelligent Control System of Ministry of Education, Taiyuan University of Technology, Taiyuan 030024, China}
\affiliation{Department of Applied Physics, The Hong Kong Polytechnic University, Hong Kong, China}

\author{Wang Tat Yau}
\affiliation{Department of Applied Physics, The Hong Kong Polytechnic University, Hong Kong, China}

\author{Yanxia Cui}
\affiliation{College of Physics and Optoelectronics, Key Lab of Advanced Transducers and Intelligent Control System of Ministry of Education, Taiyuan University of Technology, Taiyuan 030024, China}

\author{Jin Wang}
\email{jinwang@seu.edu.cn}
\affiliation{School of Physics, Southeast University, Nanjing 211189, China}

\author{Kin Hung Fung}
\email{khfung@polyu.edu.hk}
\affiliation{Department of Applied Physics, The Hong Kong Polytechnic University, Hong Kong, China}

\date{\today}

\begin{abstract}

We show that Maxwell's demon-like nonreciprocity can be supported in a class of non-Hermitian gyrotropic metasurfaces in the linear regime. The proposed metasurface functions like a transmission-only Maxwell's demon operating at a pair of photon energies. Based on the multiple scattering theory, we construct a dual-dipole model to explain the underlying mechanism that leads to the anti-symmetric nonreciprocal transmission. The results may inspire new designs of compact nonreciprocal devices for photonics.

\end{abstract}

\maketitle


Lorentz nonreciprocity in electromagnetics refers to a unique asymmetrical characteristics in the received-transmitted field ratios when the source and detector are exchanged \cite{Alu,Bloch}. Nonreciprocal elements such as isolators, gyrators, circulators or directional amplifiers have already been vital components to route signals along desired paths in the microwave systems \cite{Bernier,Peterson,Wang,Lecocq,Wang2,Sliwa}. Photonic isolation for protecting active components from backward scattering is a key requirement in stable laser devices and integrated photonic circuits \cite{Alu2,Shen,Yu,Khan,Bi}. The standard approach to break Lorentz reciprocity in the linear regime relies on the magneto-optical (MO) effect, i.e. the use of MO materials with an asymmetrical permeability (or permittivity) tensor \cite{Shal,Wong,Jin,Shi}, or parametric time-modulation composites to completely reproduce the MO effect \cite{Alu3,Lira,Sounas,Estep,Fang,Fang2}. An alternative strategy based on the nonlinear materials has also been proposed \cite{Rame,Orte,Chang,Peng,Komi,Sounas2,Sounas3,Ball}. There are respective benefits as well as limitations in different approaches \cite{Shi2,Mann}. It is believed that metamaterials or metasurfaces may offer unprecedented avenues to overcome some of the limitations.

\begin{figure}[!htbp]
\includegraphics [width=8.2cm] {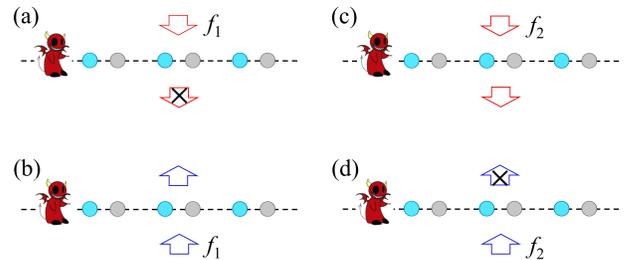}%Fig2.eps
\centering
\caption{(color online) Schematic diagram of anti-symmetric Maxwell's demon-like operation supported by a metasurface comprising of dimer unit cells. The metasurface (a) forbids the penetration of photons with frequency $f_1$ in downward incidence while it (b) allows the penetration of photons with frequency $f_1$ in upward incidence. The same metasurface also (c) allows the penetration of photons with frequency $f_2$ in downward incidence while it (d) forbids the penetration of photons with frequency $f_2$ in upward incidence.}
\label{fig0}
\end{figure}

For free-space photonics, linear elements in the form of Lorentz nonreciprocal metasurfaces are experiencing a strong surge of interest owing to their ultrathin thickness, conceptual simplicity, and potential conformability. Photonics could be a promising experimental platform for investigating the role of information in the statistical mechanics associated with Maxwell's demon \cite{vidrighin}, which is a non-reciprocal device controlling the passage of particles through a thin wall according to their energy. To date, metasurfaces supporting the Maxwell's demon-like selection of photons have not yet been addressed. In this Letter, we propose a non-Hermitian gyrotropic metasurface comprised of dimer unit cells to achieve such demon-like antisymmetric transmittance spectra for two photon energies at normal incidence. It should be noted that there were successful engineering designs supporting nonreciprocal absorption or reflection using gyromagnetic cylinders \cite{chui2010reflected,yu2012magnetically,ju2017manipulating}, but none of the mentioned designs could support demon-like action for selective transmittance. Figure \ref{fig0} schematically illustrates the Maxwell's demon-like nonreciprocity operation at two photon energies, where the demon controls nearly perfect one-way penetration for two different energy photons to particular side of metasurface, and meanwhile nearly complete rejection of reverse flow. The underlying mechanism on such demon-like nonreciprocity is explicitly revealed from the cooperative effects of magnetic rotating dipole resonance in gyromagnetic cylinders and electric dipole response in dielectric cylinders in the proposed structure. However, non-Hermiticity is found to be essential to the isolation properties, corresponding to the energy consumption of Maxwell's demon.

\begin{figure}[!htbp]
\includegraphics [width=8.2cm] {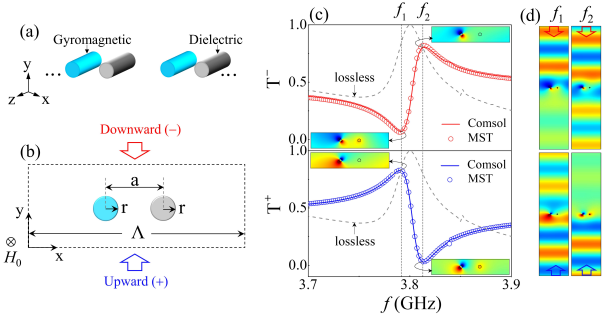}
\centering
\caption{(color online) (a) Schematic diagram of the metasurface consisting of gyromagnetic (blue) and isotropic dielectric (grey) cylinders, (b) Top view of the unit cell of the periodic system in (a). Periodic condition is applied in $x$-axis throughout all calculations. Plane wave illuminating from $-y$ and $+y$ direction are denoted as downward incidence in red arrow and upward incidence in blue arrow, respectively. (c) The numerical and analytical transmittance spectra, represented as solid line and unfilled circles respectively, under downward [top of (c)] and upward [bottom of (c)] incidences. The transmittance spectra of lossless system in both directions are shown as black dashed lines. The insets show the $\textbf{E}_z$ field patterns of those cylinders at those transmittance extrema. (d) The transient electric field $\textbf{E}_z$ at $f_1$ and $f_2$. The marked arrows indicate the direction of the incident waves. Material parameters are $\epsilon_{m}=15$, $H_0=500$ Oe, and $M_s=1750/(4\pi)$ G, and $n_d=10+2.8i$. Geometrical parameters are $\Lambda = 50$ mm, $r = 1$ mm and $a = 12$ mm.}
\label{fig1}
\end{figure}

Let us consider a metasurface as shown in Figs. \ref{fig1}(a) and \ref{fig1}(b), which is composed of an array of gyromagnetic and isotropic dielectric cylinders arranged in the $x$ direction with a lattice constant $\Lambda$. For the sake of simplicity, the gyrotropic and dielectric cylinders have the same radius $r$, and the center-to-center distance between the adjacent cylinders is denoted by $a$. A small lattice constant is chosen such that there is no diffracted propagating beams for the metasurface over the frequency range of interest. Under an external static magnetic field $H_0$ along the $-z$ axis, the chosen gyromagnetic materials, i.e. yttrium iron garnet (YIG), is described by a permittivity $\epsilon_m$ and an asymmetric permeability tensor $\tilde{\mu}$ \cite{STChui}:
\begin{equation}
\tilde{\mu}=\mu_0
\left(
\begin{tabular}{ccc}
$\mu_{1}$ & $i\mu_{2}$ & $0$\\
$-i\mu_{2}$ & $\mu_{1}$ & $0$\\
$0$ & $0$ & $1$\\
\end{tabular}
\right),
\end{equation}
with the elements $\mu_{1}=1+\omega_{m}\omega_{0}/(\omega_{0}^{2}-\omega^{2})$ and $\mu_{2}=\omega_{m}\omega/(\omega_{0}^{2}-\omega^{2})$, where $\omega_{0}=\gamma H_{0}$ is the precession frequency, $\omega_{m}=\gamma 4\pi M_{s}$, $4\pi M_{s}$ is the saturation magnetization, $\gamma$ is the gyromagnetic ratio, $\mu_0$ is the vacuum permeability, and $\omega$ is the angular frequency. The complex index of refraction of the isotropic dielectric cylinder is $n_d$, where the imaginary part is denoted as $k$. This This imaginary part contributes to energy loss which makes the dimer system non-Hermitian, resulting in nonreciprocal transmittance based on asymmetric absorption. Here, we limit the case to TE-polarized waves (electric field along the $z$ direction), and the $e^{-i\omega t}$ time-dependent convention for the harmonic field is used throughout this work.

We start with a scattering problem under the illumination of normal incident plane waves with the wavevector along downward ($-y$) direction and upward ($+y$) direction, as shown in Fig. \ref{fig1}(b). Previously, nonreciprocal absorption or reflection phenomena at oblique incidence have been demonstrated in microstructures consisting of arrays of single gyromagnetic cylinders \cite{chui2010reflected,yu2012magnetically,ju2017manipulating}. To support photonic isolation at normal incidence for Maxwell's demon-like action, additional symmetries (such as $\pi$-rotation symmetry about the $z$-axis) must be broken (which is provided by the addition of dielectric cylinders). Under the assumption of the point-dipole approximation, we may create a dual-dipole field model for the entire metasurface system by treating the gyromagnetic cylinder and the dielectric cylinder as a rotating magnetic dipole (MD) with angular quantum number -1 and damping electric dipole (ED) with angular momentum quantum number 0, respectively in the multiple scattering theory (MST) \cite{Supp}. In such dual-dipole approximation, a $2\times2$ matrix problem can be formulated and the scattering fields of the cylinders are related to the incident waves by
\begin{equation}
\label{scainc}
	\left(
		\begin{tabular}{cc}
			$s_{-1}^{Y}$ \\
			$s_{0}^{d}$ \\
		\end{tabular}
	\right)^{\mp}=
	\hat{W}^{-1}
	\left(
		\begin{tabular}{cc}
			$\mp1$ \\
			$1$ \\
		\end{tabular}
	\right)
\end{equation}
where $s_{-1}^{Y}$ and $s_{0}^{d}$ represent the scattering fields of the gyromagnetic and dielectric cylinders, respectively, and $\hat{W}^{-1}$ is an invertible response matrix in the eigen-response theory~\cite{Supp,fung}. The above equation with a sign of ``$+$" and ``$-$" represents the upward and downward incidence, respectively. The transmittance at normal incidence is then given by \cite{Supp} 
\begin{widetext}
\begin{eqnarray}
\label{T}
\text{T}^{\mp}=\left|
\frac{(L-\frac{2}{k_{0}\Lambda}+\frac{1}{b_{0}^{d}})(L-\frac{2}{k_{0}\Lambda}+\frac{1}{b_{-1}^{Y}})-(\frac{2}{k_{0}\Lambda}\pm\xi)^{2}}
{(L+\frac{1}{b_{0}^{d}})(L+\frac{1}{b_{-1}^{Y}})-\xi^{2}}
\right|^{2},
\end{eqnarray}
\end{widetext}
where $b$ is the Mie's scattering coefficient \cite{Bohren,Jin2}, and $L$ and $\xi$ are the lattice sum and the relative lattice sum \cite{Yasu,Botten} signifying the contribution of the intraspecific and interspecific coupling, respectively. The superscripts (Y/d) followed by the Mie coefficient $b$ represent the gyromagnetic/dielectric material, and the subscripts (-1/0) behind the symbols denote the angular momentum order \cite{Supp}. The nonreciprocity is triggered by the cross term $\pm4\xi/k_0\Lambda$ due to the interference of the lattice coupling $2/k_0\Lambda$ and purely imaginary interspecific coupling $\xi$ \cite{Supp}. It should be noted that the removal of absorption in dielectric cylinder (i.e $k=0$) extinguishes the nonreciprocal transmittance but the phase of transmission coefficients remains different.

The closed form in Eq. (\ref{T}) directly illustrate the anti-symmetric non-reciprocal transmittance phenomena, as indicated in Fig.~\ref{fig1}(c) with circle scatter points. Numerical results calculated by COMSOL Multiphysics are also shown in Fig.~\ref{fig1}(c) with solid lines, which shows a strong agreement with the analytical results. The transient $\textbf{E}_z$ field patterns of two tremendous transmittance difference points at frequency $f_1=3.79$ GHz [Fig.~\ref{fig1}(d), left panel] and $f_2=3.814$ GHz [Fig.~\ref{fig1}(d), right panel], where the local minimum of the spectrum in one direction nearly corresponds to the local maximum of the spectrum in opposite direction, demonstrate an imperfect Maxwell's demon function similar to Fig.~\ref{fig0}. Whatever at $f_1$ or $f_2$, the electric field strength of dielectric cylinder in low transmittance cases [Fig.~\ref{fig1}(c), Insets, right circle] is relatively higher than that in high transmittance cases.
This electric field strength may indicate the system absorption and the nonreciprocal transmittance because the loss is only introduced in dielectric cylinder.
Therefore, by removing the loss in dielectric cylinder, the transmittance spectra become identical for both directions [Fig.~\ref{fig1}(c), black dashed lines].
The  Maxwell's demon in the dimer metasurface therefore seems to be conceptually similar to statistical Maxwell's demon in the sense of the energy consumption for nonreciprocal flow. However, the demon-like action is imperfect in this system (e.g., it does not completely forbid or allow the penetration of photons in both sides \cite{Supp0}).

For two-port nonreciprocal systems, absorption loss is a necessary condition to attain nonreciprocal transmittance~\cite{Alu}.
The implement of intrinsic loss in real systems, however, affects the nonreciprocal performance in a complex way.
In many nonreciprocal designs, intrinsic loss leads to the inevitable insert loss of devices for attaining ideal nonreciprocity \cite{Khan,Dong,Dong2} while the two ports magnon photon cavity system has demonstrated the isolation in the presence of damping of modes \cite{Wang2}.
In our case, the metasurface effectively manipulates the nonreciprocity by simply controlling the absorption of dielectric cylinders. Such simple phenomenon is based on the same reflectance in both sides of the metasurface \cite{Supp} and the conservation of energy. Therefore, $\Delta \text{T}=\text{T}^--\text{T}^+=-\Delta \text{A}$, where $\Delta \text{A}=\text{A}^--\text{A}^+$ is the difference in absorption.
\begin{equation}
    \text{A}^\mp=\frac{4}{k_0\Lambda}\beta|(s_0^d)^\mp|^2
    \label{Abs}
\end{equation}
denotes the absorption in each incident direction, and $\beta=\text{Re}(L+1/b_0^d-2/k_0\Lambda)$ represents the damping coefficient.

\begin{figure}[!htbp]
\includegraphics [width=6.0cm] {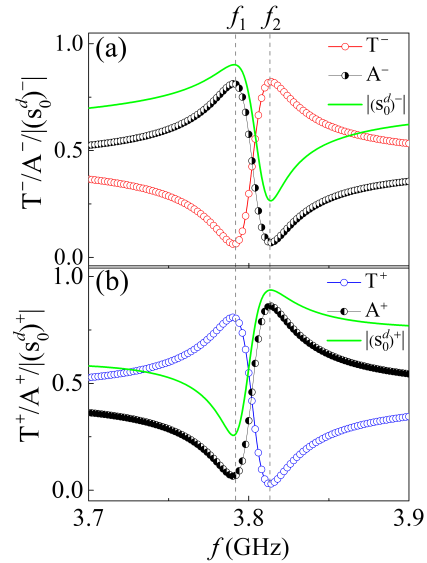}
\centering
\caption{(color online) Comparison of analytical transmittance (T) spectra (unfilled circle solid line), analytical absorption (A) spectra (half filled circle solid line) from the Eq.~(\ref{Abs}), and the magnitude of scattered field of dielectric cylinder ($|s_0^d|$) (green solid line) under (a) downward incidence and (b) upward incidence, denoted by the superscript $-$ and $+$, respectively. The local extrema of the three spectra under two incidences are simultaneously located at $f_1$ and $f_2$, marked by the black dashed lines.}
\label{fig2}
\end{figure}
\begin{figure}[!htbp]
\includegraphics [width=7.8cm] {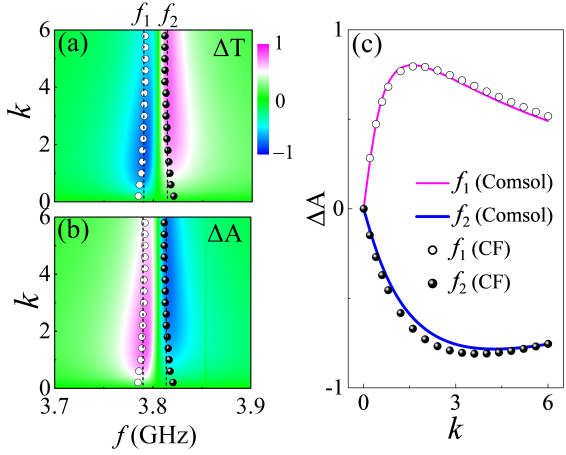}
\centering
\caption{(color online) The effects of $k$, the imaginary part of the dielectric constant, on (a)numerical transmission difference $\Delta \text{T}$ and (b) numerical absorption difference $\Delta \text{A}$. The white/black circle dots in (a) and (b) represent the peak/depth locations of difference of A in the Eq.~(\ref{Abs}). (c) The vertical cross section of (b) at $f_1$ and $f_2$, showing the effect of $k$ on $\Delta \text{A}$ at those frequencies. Excellent agreements are obtained between the closed-form (CF) in the Eq.~(\ref{Abs}) and numerical results.}
\label{ratio lss}
\end{figure}

In Fig.~\ref{fig2}, we show the relations among the transmittance ($\text{T}$) from Eq.~(\ref{T}), the absorption ($\text{A}$) in Eq.~(\ref{Abs}), and the scattered field amplitude ($s_0^d$) of dielectric cylinder. It is found that all three quantities show antisymmetric features with two opposite extrema located at $f=f_1$ and $f=f_2$, sandwiching the single-cylinder gyromagnetic resonance at $f=f_0=3.80$ GHz.
The asymmetric absorption characteristics, following the dependence of $|(s_0^d)^\mp|^2$, should therefore be the reason of triggering the nonreciprocal transmittance in the presence of $k$ (i.e., the imaginary part of dielectric constant).
As $k$ increases, the transmittance difference ($\Delta \text{T}$) [Fig.~\ref{ratio lss}(a), density plot] and absorption difference ($\Delta \text{A}$) [Fig.~\ref{ratio lss}(b), density plot] become different in signs around $f_1$ and $f_2$.
The extrema of $\Delta \text{A}$ by the Eq.~(\ref{Abs}) (Fig.~\ref{ratio lss}, white and black solid dots) also appear at $f=f_1$ and $f=f_2$.
The locations of those extrema near $f_1$ and $f_2$ are insensitive to $k$ since they are nearly pinned by the single-cylinder gyromagnetic resonance, corresponding to the frequency which satisfies $\text{Im}(1/b_{-1}^Y+L)=0$.
The $k$ dependence of the numerical and analytical values of $\Delta \text{A}$ at the two frequencies show an excellent agreement [see Fig.~\ref{ratio lss}(c)].
Therefore, the scattered field amplitude of dielectric cylinder is a good quantity to illustrate the nonreciprocity mechanism.

\begin{figure}[!htbp]
\includegraphics [width=8.2cm] {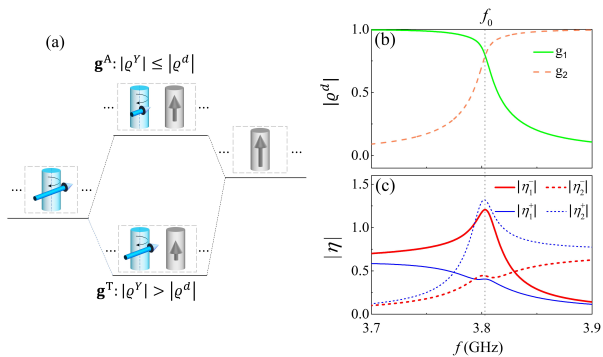}
\centering
\caption{(color online) (a) Schematic diagram of mode hybridization of gyromagnetic (blue) and dielectric (grey) cylinder array. The collective resonance mode of isolated gyromagnetic cylinder array (MD, denoted as blue arrow) and dielectric cylinder array (ED, denoted as grey arrow) are hybridized into absorptive dominant modes $\mathbf{g}^{\text{A}}$ ($|\varrho^Y|\leq |\varrho^d|$) and transmissive dominant modes $\mathbf{g}^{\text{T}}$ ($|\varrho^Y|> |\varrho^d|$) by merging two arrays. (b) The magnitudes of component corresponding to the dielectric cylinder $|\varrho^d|$ in two eigenvectors $\mathbf{g}_1$ (green lines) and $\mathbf{g}_2$ (orange dashed lines). The crossing point of two lines is found at $f_0=3.80$ GHz. (c) The magnitudes of excitation coefficient, $|\eta_{1,2}^{\pm}|$, to the two eigenvectors $\mathbf{g}_1$ and $\mathbf{g}_2$ under upward (or downward) wave excitation.
The subscripts 1, 2 of $|\eta|$ refer to the excitation of the eigenmodes with corresponding indices. The superscripts $-/+$ denote the downward/upward incidences, respectively.}
\label{fig3}
\end{figure}

To gain a deeper insight in the cooperative effect of MD in gyromagnetic cylinder and ED in dielectric one, we firstly schematically illustrate the cooperate effect in the dimer metasurface.
The isolated gyromagnetic cylinder array [Fig.~\ref{fig3}(a), blue cylinder] and the isolated dielectric cylinder array [Fig.~\ref{fig3}(a), grey cylinder] exist their own collective resonance mode.
By merging these two arrays, these two collective modes will couple and form two hybrid modes [Fig.~\ref{fig3}(a)]. In the eigen-response theory~\cite{fung}, the eigenvalues of the response matrix $\hat{{W}}^{-1}$ are given by \cite{PRA,Supp}

\begin{widetext}
\begin{equation}
1/w_{\chi}=-\frac{\frac{1}{b_0^d}+\frac{1}{b_{-1}^Y}+2L+(-1)^{\chi-1}\sqrt{(\frac{1}{b_0^d}-\frac{1}{b_{-1}^Y})^2+4\xi^2}}{2[(L+\frac{1}{b_{0}^{d}})(L+\frac{1}{b_{-1}^{Y}})-\xi^{2}]}
\end{equation}
\end{widetext}
for the corresponding eigenvector $\mathbf{g}_{\chi}$, where
\begin{equation}
\label{eig}
\mathbf{g}_{\chi}=
\left(
\begin{tabular}{cc}
$\varrho^{Y}$\\
$\varrho^{d}$\\
\end{tabular}
\right)_\chi
\end{equation}
represents the $\chi^{th}$ eigenmode of the dimer system.
$\varrho^{Y}$ and $\varrho^{d}$ in the Eq.~(\ref{eig}) denote, respectively, the components corresponding to the gyromagnetic and dielectric cylinders in the unit cell after cooperative coupling.

The normalization of eigenmodes and the birothogonality relation suggest~\cite{Supp}
\begin{equation}
    \left(
\begin{tabular}{cc}
$\varrho^{Y}$\\
$\varrho^{d}$\\
\end{tabular}
\right)_2=
\left(
\begin{tabular}{cc}
-$\varrho^{d}$\\
$\varrho^{Y}$\\
\end{tabular}
\right)_1
\end{equation}
which quantitatively classifies $\textbf{g}^\mathrm{T}$ and $\textbf{g}^\mathrm{A}$ as $\textbf{g}^\mathrm{T}:|\varrho^Y|>|\varrho^d|$ and $\textbf{g}^\mathrm{A}:|\varrho^Y|\leq|\varrho^d|$. Following the understanding of absorption and scattered field amplitude of dielectric cylinder, $\textbf{g}^\mathrm{A}$ and $\textbf{g}^\mathrm{T}$ can be realized as the loss dominant mode and transmissive dominant mode respectively.
The $|\varrho^d|$ in two eigenmodes shows the antisymmetric pairs of step function like characteristics with the transition point at $f_0\simeq 3.80$ GHz [Fig.~\ref{fig3}(b)].
The result indicates the $\textbf{g}_1$ belongs to $\textbf{g}^\mathrm{A}$ before $f_0$ and $\textbf{g}_2$ belongs to $\textbf{g}^\mathrm{A}$ after $f_0$.
As the $|\varrho^d|$ in two eigenmodes shows the large difference before or after $f_0$, the absorption seems to be dominated by $\textbf{g}^\mathbf{A}$ only and these two eigenmodes would likely be the on-off state of the gate controlled the Maxwell's demon.

To illustrate whether the gate would be opened or closed by the Maxwell's demon, we write the scattered field of cylinders through eigen-decomposition as
\begin{equation}
\label{eigsp}
\left(
\begin{tabular}{cc}
$s_{-1}^{Y}$\\
$s_{0}^{d}$\\
\end{tabular}
\right)^{\mp}
=\sum_{\chi=1}^{2}\eta_{\chi}^{\mp}\mathbf{g}_{\chi},
\label{eigd}
\end{equation}
where $\eta_\chi^\mp=(\varrho_\chi^d\mp\varrho_\chi^Y)/w_\chi$ is the complex excitation coefficient of the $\chi^{th}$ eigenmode, sensitively dependent of the downward (-) or upward (+) illumination.
The Eq.~(\ref{eigd}) shows the contribution of eigenmodes to scattered field amplitude according to their excitation coefficients.
As the absorption should be mainly contributed by the loss dominant eigenmode, the discussion of that excitation coefficient would illustrate the absorption of the system which would be interpreted as how probable for Maxwell's demon to close the gate.

Before the frequency $f_0$ where $\textbf{g}_1$ is $\textbf{g}^\mathrm{A}$, $|\eta_1^-|$ gradually increases to a maximum [Fig.~\ref{fig3}(c), red solid line] whereas $|\eta_1^+|$ drops to a local minimum [Fig.~\ref{fig3}(c), blue solid line], giving rise to the asymmetrical absorption: the rise of downward absorption and gradual drop of upward absorption.
Similarly, after $f_0$ where $\textbf{g}_2$ is $\textbf{g}^\mathrm{A}$, the distinct variation of excitation coefficients of $|\eta_2^-|$ [Fig.~\ref{fig3}(c), red dashed line] and $|\eta_2^+|$ [Fig.~\ref{fig3}(c), blue dashed line] can produce the remarkable nonreciprocity in absorption, subject to the light illumination.
As a consequence, it is revealed that the excitation of the dominant absorptive modes governs the nonreciprocal transmission, thereby the intriguing demon-like operation.
It is worthy of noting that the dimer system remains nonreciprocal by following the similar arguments, when the intrinsic material absorption is only introduced in the gyrotropic cylinder.

In summary, we propose an ultra-thin metasurface that can produce an antisymmetric nonreciprocal transmission peak-and-depth pairs. The metasurface functions like an imperfect Maxwell's demon for normal incident photons. It is found that the anti-symmetric (non-reciprocal) transmittance is induced directly by the asymmetrical absorption in lossy dielectric cylinders under the cooperative effect of the rotating magnetic dipole and linear electric dipole excited in the metasurface. Eigen-response theory using a simple $2\times2$ matrix truncated from the multiple scattering theory is employed to reveal the underlying mechanism. Moreover, the ultrathin property of the proposed metasurface is easier to realize in practical applications compared with the traditional bulky Lorentz nonreciprocal devices. Our results may inspire the designs of novel nonreciprocal devices for photonics.

W.T.Y. and W.W. contributed equally to this work. This work was supported by the Hong Kong Research Grants Council (AoE/P-02/12, C6013-18G, 15301917). W.W. and Y.C. acknowledge the support from the National Natural Science Foundation of China (61905173, 6177031853) and the Natural Science Foundation of Shanxi Province (201701D211002, 201801D221029). J.W. acknowledges the support from the Natural Science Foundation of Jiangsu Province (BK20181263).  We thank Kai Fung Lee, Wai Chun Wong, and C.T. Chan for fruitful discussion.

\bibliography{ref}

\begin{thebibliography}{48}
\expandafter\ifx\csname natexlab\endcsname\relax\def\natexlab#1{#1}\fi
\expandafter\ifx\csname bibnamefont\endcsname\relax
  \def\bibnamefont#1{#1}\fi
\expandafter\ifx\csname bibfnamefont\endcsname\relax
  \def\bibfnamefont#1{#1}\fi
\expandafter\ifx\csname citenamefont\endcsname\relax
  \def\citenamefont#1{#1}\fi
\expandafter\ifx\csname url\endcsname\relax
  \def\url#1{\texttt{#1}}\fi
\expandafter\ifx\csname urlprefix\endcsname\relax\def\urlprefix{URL }\fi
\providecommand{\bibinfo}[2]{#2}
\providecommand{\eprint}[2][]{\url{#2}}

\bibitem[{\citenamefont{Caloz et~al.}(2018)\citenamefont{Caloz, Al\`u,
  Tretyakov, Sounas, Achouri, and Deck-L\'eger}}]{Alu}
\bibinfo{author}{\bibfnamefont{C.}~\bibnamefont{Caloz}},
  \bibinfo{author}{\bibfnamefont{A.}~\bibnamefont{Al\`u}},
  \bibinfo{author}{\bibfnamefont{S.}~\bibnamefont{Tretyakov}},
  \bibinfo{author}{\bibfnamefont{D.}~\bibnamefont{Sounas}},
  \bibinfo{author}{\bibfnamefont{K.}~\bibnamefont{Achouri}}, \bibnamefont{and}
  \bibinfo{author}{\bibfnamefont{Z.-L.} \bibnamefont{Deck-L\'eger}},
\bibinfo {title} {Electromagnetic nonreciprocity},
  \bibinfo{journal}{Phys. Rev. Applied} \textbf{\bibinfo{volume}{10}},
  \bibinfo{pages}{047001} (\bibinfo{year}{2018}).

\bibitem[{\citenamefont{Rodriguez et~al.}(2019)\citenamefont{Rodriguez, Goblot,
  Zambon, Amo, and Bloch}}]{Bloch}
\bibinfo{author}{\bibfnamefont{S.~R.~K.} \bibnamefont{Rodriguez}},
  \bibinfo{author}{\bibfnamefont{V.}~\bibnamefont{Goblot}},
  \bibinfo{author}{\bibfnamefont{N.~C.} \bibnamefont{Zambon}},
  \bibinfo{author}{\bibfnamefont{A.}~\bibnamefont{Amo}}, \bibnamefont{and}
  \bibinfo{author}{\bibfnamefont{J.}~\bibnamefont{Bloch}},
\bibinfo {title} {Nonreciprocity and zero reflection in nonlinear cavities with tailored loss},
  \bibinfo{journal}{Phys. Rev. A} \textbf{\bibinfo{volume}{99}},
  \bibinfo{pages}{013851} (\bibinfo{year}{2019}).

\bibitem[{\citenamefont{Bernier et~al.}(2017)\citenamefont{Bernier, Toth,
  Koottandavida, Ioannou, Malz, Nunnenkamp, Feofanov, and
  Kippenberg}}]{Bernier}
\bibinfo{author}{\bibfnamefont{N.~R.} \bibnamefont{Bernier}},
  \bibinfo{author}{\bibfnamefont{L.~D.} \bibnamefont{Toth}},
  \bibinfo{author}{\bibfnamefont{A.}~\bibnamefont{Koottandavida}},
  \bibinfo{author}{\bibfnamefont{M.~A.} \bibnamefont{Ioannou}},
  \bibinfo{author}{\bibfnamefont{D.}~\bibnamefont{Malz}},
  \bibinfo{author}{\bibfnamefont{A.}~\bibnamefont{Nunnenkamp}},
  \bibinfo{author}{\bibfnamefont{A.}~\bibnamefont{Feofanov}}, \bibnamefont{and}
  \bibinfo{author}{\bibfnamefont{T.}~\bibnamefont{Kippenberg}},
\bibinfo {title} {Nonreciprocal reconfigurable microwave optomechanical circuit},
  \bibinfo{journal}{Nature communications} \textbf{\bibinfo{volume}{8}},
  \bibinfo{pages}{1} (\bibinfo{year}{2017}).

\bibitem[{\citenamefont{Peterson et~al.}(2017)\citenamefont{Peterson, Lecocq,
  Cicak, Simmonds, Aumentado, and Teufel}}]{Peterson}
\bibinfo{author}{\bibfnamefont{G.~A.} \bibnamefont{Peterson}},
  \bibinfo{author}{\bibfnamefont{F.}~\bibnamefont{Lecocq}},
  \bibinfo{author}{\bibfnamefont{K.}~\bibnamefont{Cicak}},
  \bibinfo{author}{\bibfnamefont{R.~W.} \bibnamefont{Simmonds}},
  \bibinfo{author}{\bibfnamefont{J.}~\bibnamefont{Aumentado}},
  \bibnamefont{and} \bibinfo{author}{\bibfnamefont{J.~D.}
  \bibnamefont{Teufel}}, \bibinfo{journal}{Phys. Rev. X}

\bibitem[{\citenamefont{Wang et~al.}(2009)\citenamefont{Wang, Chong,
  Joannopoulos, and Solja{\v{c}}i{\'c}}}]{Wang}
\bibinfo{author}{\bibfnamefont{Z.}~\bibnamefont{Wang}},
  \bibinfo{author}{\bibfnamefont{Y.}~\bibnamefont{Chong}},
  \bibinfo{author}{\bibfnamefont{J.~D.} \bibnamefont{Joannopoulos}},
  \bibnamefont{and}
  \bibinfo{author}{\bibfnamefont{M.}~\bibnamefont{Solja{\v{c}}i{\'c}}},
\bibinfo {title} {Demonstration of efficient nonreciprocity in a microwave optomechanical circuit},  \textbf{\bibinfo{volume}{7}}, \bibinfo{pages}{031001} (\bibinfo{year}{2017}).
  \bibinfo{journal}{Nature} \textbf{\bibinfo{volume}{461}},
  \bibinfo{pages}{772} (\bibinfo{year}{2009}).

\bibitem[{\citenamefont{Lecocq et~al.}(2017)\citenamefont{Lecocq, Ranzani,
  Peterson, Cicak, Simmonds, Teufel, and Aumentado}}]{Lecocq}
\bibinfo{author}{\bibfnamefont{F.}~\bibnamefont{Lecocq}},
  \bibinfo{author}{\bibfnamefont{L.}~\bibnamefont{Ranzani}},
  \bibinfo{author}{\bibfnamefont{G.~A.} \bibnamefont{Peterson}},
  \bibinfo{author}{\bibfnamefont{K.}~\bibnamefont{Cicak}},
  \bibinfo{author}{\bibfnamefont{R.~W.} \bibnamefont{Simmonds}},
  \bibinfo{author}{\bibfnamefont{J.~D.} \bibnamefont{Teufel}},
  \bibnamefont{and}
  \bibinfo{author}{\bibfnamefont{J.}~\bibnamefont{Aumentado}},
\bibinfo {title} {Nonreciprocal microwave signal processing with a field-programmable josephson amplifier},
  \bibinfo{journal}{Phys. Rev. Applied} \textbf{\bibinfo{volume}{7}},
  \bibinfo{pages}{024028} (\bibinfo{year}{2017}).

\bibitem[{\citenamefont{Wang et~al.}(2019)\citenamefont{Wang, Rao, Yang, Xu,
  Gui, Yao, You, and Hu}}]{Wang2}
\bibinfo{author}{\bibfnamefont{Y.-P.} \bibnamefont{Wang}},
  \bibinfo{author}{\bibfnamefont{J.~W.} \bibnamefont{Rao}},
  \bibinfo{author}{\bibfnamefont{Y.}~\bibnamefont{Yang}},
  \bibinfo{author}{\bibfnamefont{P.-C.} \bibnamefont{Xu}},
  \bibinfo{author}{\bibfnamefont{Y.~S.} \bibnamefont{Gui}},
  \bibinfo{author}{\bibfnamefont{B.~M.} \bibnamefont{Yao}},
  \bibinfo{author}{\bibfnamefont{J.~Q.} \bibnamefont{You}}, \bibnamefont{and}
  \bibinfo{author}{\bibfnamefont{C.-M.} \bibnamefont{Hu}},
\bibinfo {title}{Nonreciprocity and unidirectional invisibility in cavity magnonics},
  \bibinfo{journal}{Phys. Rev. Lett.} \textbf{\bibinfo{volume}{123}},
  \bibinfo{pages}{127202} (\bibinfo{year}{2019}).

\bibitem[{\citenamefont{Sliwa et~al.}(2015)\citenamefont{Sliwa, Hatridge,
  Narla, Shankar, Frunzio, Schoelkopf, and Devoret}}]{Sliwa}
\bibinfo{author}{\bibfnamefont{K.~M.} \bibnamefont{Sliwa}},
  \bibinfo{author}{\bibfnamefont{M.}~\bibnamefont{Hatridge}},
  \bibinfo{author}{\bibfnamefont{A.}~\bibnamefont{Narla}},
  \bibinfo{author}{\bibfnamefont{S.}~\bibnamefont{Shankar}},
  \bibinfo{author}{\bibfnamefont{L.}~\bibnamefont{Frunzio}},
  \bibinfo{author}{\bibfnamefont{R.~J.} \bibnamefont{Schoelkopf}},
  \bibnamefont{and} \bibinfo{author}{\bibfnamefont{M.~H.}
  \bibnamefont{Devoret}}, 
\bibinfo {title} {Reconfigurable josephson circulator/directional amplifier},
  \bibinfo{journal}{Phys. Rev. X}
  \textbf{\bibinfo{volume}{5}}, \bibinfo{pages}{041020} (\bibinfo{year}{2015})

\bibitem[{\citenamefont{Sounas and Al{\`u}}(2017{\natexlab{a}})}]{Alu2}
\bibinfo{author}{\bibfnamefont{D.~L.} \bibnamefont{Sounas}} \bibnamefont{and}
  \bibinfo{author}{\bibfnamefont{A.}~\bibnamefont{Al{\`u}}},
\bibinfo {title}{Non-reciprocal photonics based on time modulation},
  \bibinfo{journal}{Nature Photonics} \textbf{\bibinfo{volume}{11}},
  \bibinfo{pages}{774} (\bibinfo{year}{2017}{\natexlab{a}}).

\bibitem[{\citenamefont{Shen et~al.}(2016)\citenamefont{Shen, Zhang, Chen, Zou,
  Xiao, Zou, Sun, Guo, and Dong}}]{Shen}
\bibinfo{author}{\bibfnamefont{Z.}~\bibnamefont{Shen}},
  \bibinfo{author}{\bibfnamefont{Y.-L.} \bibnamefont{Zhang}},
  \bibinfo{author}{\bibfnamefont{Y.}~\bibnamefont{Chen}},
  \bibinfo{author}{\bibfnamefont{C.-L.} \bibnamefont{Zou}},
  \bibinfo{author}{\bibfnamefont{Y.-F.} \bibnamefont{Xiao}},
  \bibinfo{author}{\bibfnamefont{X.-B.} \bibnamefont{Zou}},
  \bibinfo{author}{\bibfnamefont{F.-W.} \bibnamefont{Sun}},
  \bibinfo{author}{\bibfnamefont{G.-C.} \bibnamefont{Guo}}, \bibnamefont{and}
  \bibinfo{author}{\bibfnamefont{C.-H.} \bibnamefont{Dong}},
\bibinfo {title} {Experimental realization of optomechanically induced non-reciprocity},
  \bibinfo{journal}{Nature Photonics} \textbf{\bibinfo{volume}{10}},
  \bibinfo{pages}{657} (\bibinfo{year}{2016}).

\bibitem[{\citenamefont{Yu and Fan}(2009)}]{Yu}
\bibinfo{author}{\bibfnamefont{Z.}~\bibnamefont{Yu}} \bibnamefont{and}
  \bibinfo{author}{\bibfnamefont{S.}~\bibnamefont{Fan}},
\bibinfo {title} {Complete optical isolation created by indirect interband photonic transitions},
  \bibinfo{journal}{Nature photonics} \textbf{\bibinfo{volume}{3}},
  \bibinfo{pages}{91} (\bibinfo{year}{2009}).

\bibitem[{\citenamefont{Khanikaev et~al.}(2010)\citenamefont{Khanikaev,
  Mousavi, Shvets, and Kivshar}}]{Khan}
\bibinfo{author}{\bibfnamefont{A.~B.} \bibnamefont{Khanikaev}},
  \bibinfo{author}{\bibfnamefont{S.~H.} \bibnamefont{Mousavi}},
  \bibinfo{author}{\bibfnamefont{G.}~\bibnamefont{Shvets}}, \bibnamefont{and}
  \bibinfo{author}{\bibfnamefont{Y.~S.} \bibnamefont{Kivshar}},
\bibinfo {title}{One-way extraordinary optical transmission and nonreciprocal spoof plasmons},
  \bibinfo{journal}{Phys. Rev. Lett.} \textbf{\bibinfo{volume}{105}},
  \bibinfo{pages}{126804} (\bibinfo{year}{2010}).

\bibitem[{\citenamefont{Bi et~al.}(2011)\citenamefont{Bi, Hu, Jiang, Kim,
  Dionne, Kimerling, and Ross}}]{Bi}
\bibinfo{author}{\bibfnamefont{L.}~\bibnamefont{Bi}},
  \bibinfo{author}{\bibfnamefont{J.}~\bibnamefont{Hu}},
  \bibinfo{author}{\bibfnamefont{P.}~\bibnamefont{Jiang}},
  \bibinfo{author}{\bibfnamefont{D.~H.} \bibnamefont{Kim}},
  \bibinfo{author}{\bibfnamefont{G.~F.} \bibnamefont{Dionne}},
  \bibinfo{author}{\bibfnamefont{L.~C.} \bibnamefont{Kimerling}},
  \bibnamefont{and} \bibinfo{author}{\bibfnamefont{C.}~\bibnamefont{Ross}},
\bibinfo {title}{On-chip optical isolation in monolithically integrated non-reciprocal optical resonators},
  \bibinfo{journal}{Nature Photonics} \textbf{\bibinfo{volume}{5}},
  \bibinfo{pages}{758} (\bibinfo{year}{2011}).

\bibitem[{\citenamefont{Shalaby et~al.}(2013)\citenamefont{Shalaby, Peccianti,
  Ozturk, and Morandotti}}]{Shal}
\bibinfo{author}{\bibfnamefont{M.}~\bibnamefont{Shalaby}},
  \bibinfo{author}{\bibfnamefont{M.}~\bibnamefont{Peccianti}},
  \bibinfo{author}{\bibfnamefont{Y.}~\bibnamefont{Ozturk}}, \bibnamefont{and}
  \bibinfo{author}{\bibfnamefont{R.}~\bibnamefont{Morandotti}},
\bibinfo {title} {A magnetic non-reciprocal isolator for broadband terahertz operation},
  \bibinfo{journal}{Nature communications} \textbf{\bibinfo{volume}{4}},
  \bibinfo{pages}{1} (\bibinfo{year}{2013}).

\bibitem[{\citenamefont{Wong et~al.}(2020)\citenamefont{Wong, Wang, Yau, and
  Fung}}]{Wong}
\bibinfo{author}{\bibfnamefont{W.~C.} \bibnamefont{Wong}},
  \bibinfo{author}{\bibfnamefont{W.}~\bibnamefont{Wang}},
  \bibinfo{author}{\bibfnamefont{W.~T.} \bibnamefont{Yau}}, \bibnamefont{and}
  \bibinfo{author}{\bibfnamefont{K.~H.} \bibnamefont{Fung}},
\bibinfo {title} {Topological theory for perfect metasurface isolators},
  \bibinfo{journal}{Phys. Rev. B} \textbf{\bibinfo{volume}{101}},
  \bibinfo{pages}{121405} (\bibinfo{year}{2020}).

\bibitem[{\citenamefont{Wang et~al.}(2018)\citenamefont{Wang, Dong, Shi, Wang,
  and Fung}}]{Jin}
\bibinfo{author}{\bibfnamefont{J.}~\bibnamefont{Wang}},
  \bibinfo{author}{\bibfnamefont{H.~Y.} \bibnamefont{Dong}},
  \bibinfo{author}{\bibfnamefont{Q.~Y.} \bibnamefont{Shi}},
  \bibinfo{author}{\bibfnamefont{W.}~\bibnamefont{Wang}}, \bibnamefont{and}
  \bibinfo{author}{\bibfnamefont{K.~H.} \bibnamefont{Fung}},
\bibinfo {title}{Coalescence of nonreciprocal exceptional points in magnetized \textsl{PT}-symmetric systems},
  \bibinfo{journal}{Phys. Rev. B} \textbf{\bibinfo{volume}{97}},
  \bibinfo{pages}{014428} (\bibinfo{year}{2018}).

\bibitem[{\citenamefont{Shi et~al.}(2018)\citenamefont{Shi, Dong, Fung, gao
  Dong, and Wang}}]{Shi}
\bibinfo{author}{\bibfnamefont{Q.~Y.} \bibnamefont{Shi}},
  \bibinfo{author}{\bibfnamefont{H.~Y.} \bibnamefont{Dong}},
  \bibinfo{author}{\bibfnamefont{K.~H.} \bibnamefont{Fung}},
  \bibinfo{author}{\bibfnamefont{Z.}~\bibnamefont{gao Dong}}, \bibnamefont{and}
  \bibinfo{author}{\bibfnamefont{J.}~\bibnamefont{Wang}},
\bibinfo {title} {Optical non-reciprocity induced by asymmetrical dispersion of tamm plasmon polaritons in terahertz magnetoplasmonic crystals},
  \bibinfo{journal}{Opt. Express} \textbf{\bibinfo{volume}{26}},
  \bibinfo{pages}{33613} (\bibinfo{year}{2018}).

\bibitem[{\citenamefont{Sounas and Al{\`u}}(2017{\natexlab{b}})}]{Alu3}
\bibinfo{author}{\bibfnamefont{D.~L.} \bibnamefont{Sounas}} \bibnamefont{and}
  \bibinfo{author}{\bibfnamefont{A.}~\bibnamefont{Al{\`u}}},
\bibinfo {title}{Non-reciprocal photonics based on time modulation},
  \bibinfo{journal}{Nature Photonics} \textbf{\bibinfo{volume}{11}},
  \bibinfo{pages}{774} (\bibinfo{year}{2017}{\natexlab{b}}).

\bibitem[{\citenamefont{Lira et~al.}(2012)\citenamefont{Lira, Yu, Fan, and
  Lipson}}]{Lira}
\bibinfo{author}{\bibfnamefont{H.}~\bibnamefont{Lira}},
  \bibinfo{author}{\bibfnamefont{Z.}~\bibnamefont{Yu}},
  \bibinfo{author}{\bibfnamefont{S.}~\bibnamefont{Fan}}, \bibnamefont{and}
  \bibinfo{author}{\bibfnamefont{M.}~\bibnamefont{Lipson}},
\bibinfo {title}{Electrically driven nonreciprocity induced by interband photonic transition on a silicon chip},
  \bibinfo{journal}{Phys. Rev. Lett.} \textbf{\bibinfo{volume}{109}},
  \bibinfo{pages}{033901} (\bibinfo{year}{2012}).

\bibitem[{\citenamefont{Sounas et~al.}(2013)\citenamefont{Sounas, Caloz, and
  Alu}}]{Sounas}
\bibinfo{author}{\bibfnamefont{D.~L.} \bibnamefont{Sounas}},
  \bibinfo{author}{\bibfnamefont{C.}~\bibnamefont{Caloz}}, \bibnamefont{and}
  \bibinfo{author}{\bibfnamefont{A.}~\bibnamefont{Alu}},
\bibinfo {title} {Giant non-reciprocity at the subwavelength scale using angular momentum-biased metamaterials},
  \bibinfo{journal}{Nature communications} \textbf{\bibinfo{volume}{4}},
  \bibinfo{pages}{2407} (\bibinfo{year}{2013}).

\bibitem[{\citenamefont{Estep et~al.}(2014)\citenamefont{Estep, Sounas, Soric,
  and Al{\`u}}}]{Estep}
\bibinfo{author}{\bibfnamefont{N.~A.} \bibnamefont{Estep}},
  \bibinfo{author}{\bibfnamefont{D.~L.} \bibnamefont{Sounas}},
  \bibinfo{author}{\bibfnamefont{J.}~\bibnamefont{Soric}}, \bibnamefont{and}
  \bibinfo{author}{\bibfnamefont{A.}~\bibnamefont{Al{\`u}}},
\bibinfo {title}{Magnetic-free non-reciprocity and isolation based on parametrically modulated coupled-resonator loops},
  \bibinfo{journal}{Nature Physics} \textbf{\bibinfo{volume}{10}},
  \bibinfo{pages}{923} (\bibinfo{year}{2014}).

\bibitem[{\citenamefont{Fang et~al.}(2012{\natexlab{a}})\citenamefont{Fang, Yu,
  and Fan}}]{Fang}
\bibinfo{author}{\bibfnamefont{K.}~\bibnamefont{Fang}},
  \bibinfo{author}{\bibfnamefont{Z.}~\bibnamefont{Yu}}, \bibnamefont{and}
  \bibinfo{author}{\bibfnamefont{S.}~\bibnamefont{Fan}},
\bibinfo {title} {Photonic aharonov-bohm effect based on dynamic modulation},
  \bibinfo{journal}{Phys. Rev. Lett.} \textbf{\bibinfo{volume}{108}},
  \bibinfo{pages}{153901} (\bibinfo{year}{2012}{\natexlab{a}}).

\bibitem[{\citenamefont{Fang et~al.}(2012{\natexlab{b}})\citenamefont{Fang, Yu,
  and Fan}}]{Fang2}
\bibinfo{author}{\bibfnamefont{K.}~\bibnamefont{Fang}},
  \bibinfo{author}{\bibfnamefont{Z.}~\bibnamefont{Yu}}, \bibnamefont{and}
  \bibinfo{author}{\bibfnamefont{S.}~\bibnamefont{Fan}},
\bibinfo {title} {Realizing effective magnetic field for photons by controlling the phase of dynamic modulation},
  \bibinfo{journal}{Nature photonics} \textbf{\bibinfo{volume}{6}},
  \bibinfo{pages}{782} (\bibinfo{year}{2012}{\natexlab{b}}).

\bibitem[{\citenamefont{Ramezani et~al.}(2010)\citenamefont{Ramezani, Kottos,
  El-Ganainy, and Christodoulides}}]{Rame}
\bibinfo{author}{\bibfnamefont{H.}~\bibnamefont{Ramezani}},
  \bibinfo{author}{\bibfnamefont{T.}~\bibnamefont{Kottos}},
  \bibinfo{author}{\bibfnamefont{R.}~\bibnamefont{El-Ganainy}},
  \bibnamefont{and} \bibinfo{author}{\bibfnamefont{D.~N.}
  \bibnamefont{Christodoulides}},
\bibinfo {title} {Unidirectional nonlinear \textsl{PT}-symmetric optical structures},
 \bibinfo{journal}{Phys. Rev. A}
  \textbf{\bibinfo{volume}{82}}, \bibinfo{pages}{043803}
  (\bibinfo{year}{2010}).

\bibitem[{\citenamefont{Espinosa-Ortega
  et~al.}(2013)\citenamefont{Espinosa-Ortega, Liew, and Shelykh}}]{Orte}
\bibinfo{author}{\bibfnamefont{T.}~\bibnamefont{Espinosa-Ortega}},
  \bibinfo{author}{\bibfnamefont{T.~C.~H.} \bibnamefont{Liew}},
  \bibnamefont{and} \bibinfo{author}{\bibfnamefont{I.~A.}
  \bibnamefont{Shelykh}},
\bibinfo {title}{Optical diode based on exciton-polaritons},
 \bibinfo{journal}{Applied Physics Letters}
  \textbf{\bibinfo{volume}{103}}, \bibinfo{pages}{191110}
  (\bibinfo{year}{2013}).

\bibitem[{\citenamefont{Chang et~al.}(2014)\citenamefont{Chang, Jiang, Hua,
  Yang, Wen, Jiang, Li, Wang, and Xiao}}]{Chang}
\bibinfo{author}{\bibfnamefont{L.}~\bibnamefont{Chang}},
  \bibinfo{author}{\bibfnamefont{X.}~\bibnamefont{Jiang}},
  \bibinfo{author}{\bibfnamefont{S.}~\bibnamefont{Hua}},
  \bibinfo{author}{\bibfnamefont{C.}~\bibnamefont{Yang}},
  \bibinfo{author}{\bibfnamefont{J.}~\bibnamefont{Wen}},
  \bibinfo{author}{\bibfnamefont{L.}~\bibnamefont{Jiang}},
  \bibinfo{author}{\bibfnamefont{G.}~\bibnamefont{Li}},
  \bibinfo{author}{\bibfnamefont{G.}~\bibnamefont{Wang}}, \bibnamefont{and}
  \bibinfo{author}{\bibfnamefont{M.}~\bibnamefont{Xiao}},
\bibinfo {title}{Parity--time symmetry and variable optical isolation in active--passive-coupled microresonators},
  \bibinfo{journal}{Nature photonics} \textbf{\bibinfo{volume}{8}},
  \bibinfo{pages}{524} (\bibinfo{year}{2014}).

\bibitem[{\citenamefont{Peng et~al.}(2014)\citenamefont{Peng, {\"O}zdemir, Lei,
  Monifi, Gianfreda, Long, Fan, Nori, Bender, and Yang}}]{Peng}
\bibinfo{author}{\bibfnamefont{B.}~\bibnamefont{Peng}},
  \bibinfo{author}{\bibfnamefont{{\c{S}}.~K.} \bibnamefont{{\"O}zdemir}},
  \bibinfo{author}{\bibfnamefont{F.}~\bibnamefont{Lei}},
  \bibinfo{author}{\bibfnamefont{F.}~\bibnamefont{Monifi}},
  \bibinfo{author}{\bibfnamefont{M.}~\bibnamefont{Gianfreda}},
  \bibinfo{author}{\bibfnamefont{G.~L.} \bibnamefont{Long}},
  \bibinfo{author}{\bibfnamefont{S.}~\bibnamefont{Fan}},
  \bibinfo{author}{\bibfnamefont{F.}~\bibnamefont{Nori}},
  \bibinfo{author}{\bibfnamefont{C.~M.} \bibnamefont{Bender}},
  \bibnamefont{and} \bibinfo{author}{\bibfnamefont{L.}~\bibnamefont{Yang}},
\bibinfo {title} {Parity--time-symmetric whispering-gallery microcavities},
  \bibinfo{journal}{Nature Physics} \textbf{\bibinfo{volume}{10}},
  \bibinfo{pages}{394} (\bibinfo{year}{2014}).

\bibitem[{\citenamefont{Kominis et~al.}(2016)\citenamefont{Kominis, Bountis,
  and Flach}}]{Komi}
\bibinfo{author}{\bibfnamefont{Y.}~\bibnamefont{Kominis}},
  \bibinfo{author}{\bibfnamefont{T.}~\bibnamefont{Bountis}}, \bibnamefont{and}
  \bibinfo{author}{\bibfnamefont{S.}~\bibnamefont{Flach}},
\bibinfo {title}{The asymmetric active coupler: Stable nonlinear supermodes and directed transport},
  \bibinfo{journal}{Scientific reports} \textbf{\bibinfo{volume}{6}},
  \bibinfo{pages}{1} (\bibinfo{year}{2016}).

\bibitem[{\citenamefont{Sounas and Al\`u}(2017)}]{Sounas2}
\bibinfo{author}{\bibfnamefont{D.~L.} \bibnamefont{Sounas}} \bibnamefont{and}
  \bibinfo{author}{\bibfnamefont{A.}~\bibnamefont{Al\`u}},
\bibinfo {title}{Time-reversal symmetry bounds on the electromagnetic response of asymmetric structures},
  \bibinfo{journal}{Phys. Rev. Lett.} \textbf{\bibinfo{volume}{118}},
  \bibinfo{pages}{154302} (\bibinfo{year}{2017}).

\bibitem[{\citenamefont{Sounas et~al.}(2018)\citenamefont{Sounas, Soric, and
  Alu}}]{Sounas3}
\bibinfo{author}{\bibfnamefont{D.~L.} \bibnamefont{Sounas}},
  \bibinfo{author}{\bibfnamefont{J.}~\bibnamefont{Soric}}, \bibnamefont{and}
  \bibinfo{author}{\bibfnamefont{A.}~\bibnamefont{Alu}},
\bibinfo {title}{Broadband passive isolators based on coupled nonlinear resonances},
  \bibinfo{journal}{Nature Electronics} \textbf{\bibinfo{volume}{1}},
  \bibinfo{pages}{113} (\bibinfo{year}{2018}).

\bibitem[{\citenamefont{Ballarini et~al.}(2013)\citenamefont{Ballarini,
  De~Giorgi, Cancellieri, Houdr{\'e}, Giacobino, Cingolani, Bramati, Gigli, and
  Sanvitto}}]{Ball}
\bibinfo{author}{\bibfnamefont{D.}~\bibnamefont{Ballarini}},
  \bibinfo{author}{\bibfnamefont{M.}~\bibnamefont{De~Giorgi}},
  \bibinfo{author}{\bibfnamefont{E.}~\bibnamefont{Cancellieri}},
  \bibinfo{author}{\bibfnamefont{R.}~\bibnamefont{Houdr{\'e}}},
  \bibinfo{author}{\bibfnamefont{E.}~\bibnamefont{Giacobino}},
  \bibinfo{author}{\bibfnamefont{R.}~\bibnamefont{Cingolani}},
  \bibinfo{author}{\bibfnamefont{A.}~\bibnamefont{Bramati}},
  \bibinfo{author}{\bibfnamefont{G.}~\bibnamefont{Gigli}}, \bibnamefont{and}
  \bibinfo{author}{\bibfnamefont{D.}~\bibnamefont{Sanvitto}},
\bibinfo {title}{All-optical polariton transistor},
  \bibinfo{journal}{Nature communications} \textbf{\bibinfo{volume}{4}},
  \bibinfo{pages}{1} (\bibinfo{year}{2013}).

\bibitem[{\citenamefont{Shi et~al.}(2015)\citenamefont{Shi, Yu, and
  Fan}}]{Shi2}
\bibinfo{author}{\bibfnamefont{Y.}~\bibnamefont{Shi}},
  \bibinfo{author}{\bibfnamefont{Z.}~\bibnamefont{Yu}}, \bibnamefont{and}
  \bibinfo{author}{\bibfnamefont{S.}~\bibnamefont{Fan}},
\bibinfo {title} {Limitations of nonlinear optical isolators due to dynamic reciprocity},
  \bibinfo{journal}{Nature photonics} \textbf{\bibinfo{volume}{9}},
  \bibinfo{pages}{388} (\bibinfo{year}{2015}).

\bibitem[{\citenamefont{Mann et~al.}(2019)\citenamefont{Mann, Sounas, and
  Al\`{u}}}]{Mann}
\bibinfo{author}{\bibfnamefont{S.~A.} \bibnamefont{Mann}},
  \bibinfo{author}{\bibfnamefont{D.~L.} \bibnamefont{Sounas}},
  \bibnamefont{and} \bibinfo{author}{\bibfnamefont{A.}~\bibnamefont{Al\`{u}}},
\bibinfo {title}{Nonreciprocal cavities and the time--bandwidth limit},
  \bibinfo{journal}{Optica} \textbf{\bibinfo{volume}{6}}, \bibinfo{pages}{104}
  (\bibinfo{year}{2019}).

\bibitem[{\citenamefont{Vidrighin et~al.}(2016)\citenamefont{Vidrighin,
  Dahlsten, Barbieri, Kim, Vedral, and Walmsley}}]{vidrighin}
\bibinfo{author}{\bibfnamefont{M.~D.} \bibnamefont{Vidrighin}},
  \bibinfo{author}{\bibfnamefont{O.}~\bibnamefont{Dahlsten}},
  \bibinfo{author}{\bibfnamefont{M.}~\bibnamefont{Barbieri}},
  \bibinfo{author}{\bibfnamefont{M.~S.} \bibnamefont{Kim}},
  \bibinfo{author}{\bibfnamefont{V.}~\bibnamefont{Vedral}}, \bibnamefont{and}
  \bibinfo{author}{\bibfnamefont{I.~A.} \bibnamefont{Walmsley}},
\bibinfo {title} {Photonic maxwell's demon},
  \bibinfo{journal}{Phys. Rev. Lett.} \textbf{\bibinfo{volume}{116}},
  \bibinfo{pages}{050401} (\bibinfo{year}{2016}).

\bibitem[{\citenamefont{Chui et~al.}(2010)\citenamefont{Chui, Liu, and
  Lin}}]{chui2010reflected}
\bibinfo{author}{\bibfnamefont{S.}~\bibnamefont{Chui}},
  \bibinfo{author}{\bibfnamefont{S.}~\bibnamefont{Liu}}, \bibnamefont{and}
  \bibinfo{author}{\bibfnamefont{Z.}~\bibnamefont{Lin}},
\bibinfo {title} {Reflected wave of finite circulation from magnetic photonic crystals},
  \bibinfo{journal}{Journal of Physics: Condensed Matter}
  \textbf{\bibinfo{volume}{22}}, \bibinfo{pages}{182201}
  (\bibinfo{year}{2010}).

\bibitem[{\citenamefont{Yu et~al.}(2012)\citenamefont{Yu, Chen, Wu, and
  Liu}}]{yu2012magnetically}
\bibinfo{author}{\bibfnamefont{J.}~\bibnamefont{Yu}},
  \bibinfo{author}{\bibfnamefont{H.}~\bibnamefont{Chen}},
  \bibinfo{author}{\bibfnamefont{Y.}~\bibnamefont{Wu}}, \bibnamefont{and}
  \bibinfo{author}{\bibfnamefont{S.}~\bibnamefont{Liu}},
\bibinfo {title} {Magnetically manipulable perfect unidirectional absorber based on nonreciprocal magnetic surface plasmon},
 \bibinfo{journal}{EPL
  (Europhysics Letters)} \textbf{\bibinfo{volume}{100}}, \bibinfo{pages}{47007}
  (\bibinfo{year}{2012}).

\bibitem[{\citenamefont{Ju et~al.}(2017)\citenamefont{Ju, Wu, Li, Poo, Liu, and
  Lin}}]{ju2017manipulating}
\bibinfo{author}{\bibfnamefont{C.}~\bibnamefont{Ju}},
  \bibinfo{author}{\bibfnamefont{R.-X.} \bibnamefont{Wu}},
  \bibinfo{author}{\bibfnamefont{Z.}~\bibnamefont{Li}},
  \bibinfo{author}{\bibfnamefont{Y.}~\bibnamefont{Poo}},
  \bibinfo{author}{\bibfnamefont{S.-Y.} \bibnamefont{Liu}}, \bibnamefont{and}
  \bibinfo{author}{\bibfnamefont{Z.-F.} \bibnamefont{Lin}},
\bibinfo {title} {Manipulating electromagnetic wave propagating non-reciprocally by a chain of ferrite rods},
  \bibinfo{journal}{Optics Express} \textbf{\bibinfo{volume}{25}},
  \bibinfo{pages}{22096} (\bibinfo{year}{2017}).

\bibitem[{\citenamefont{Jin et~al.}(2009)\citenamefont{Jin, Liu, Lin, and
  Chui}}]{STChui}
\bibinfo{author}{\bibfnamefont{J.}~\bibnamefont{Jin}},
  \bibinfo{author}{\bibfnamefont{S.}~\bibnamefont{Liu}},
  \bibinfo{author}{\bibfnamefont{Z.}~\bibnamefont{Lin}}, \bibnamefont{and}
  \bibinfo{author}{\bibfnamefont{S.~T.} \bibnamefont{Chui}},
\bibinfo {title} {Effective-medium theory for anisotropic magnetic metamaterials},
  \bibinfo{journal}{Phys. Rev. B} \textbf{\bibinfo{volume}{80}},
  \bibinfo{pages}{115101} (\bibinfo{year}{2009}).

\bibitem[{Sup({\natexlab{a}})}]{Supp}
\bibinfo{note}{See Supplemental Material at [URL will be inserted by publisher]
  for the derivation of dual dipole model from Multiple Scattering theory and
  spectra analysis}.

\bibitem[{\citenamefont{Fung et~al.}(2014)\citenamefont{Fung, Kumar, and
  Fang}}]{fung}
\bibinfo{author}{\bibfnamefont{K.~H.} \bibnamefont{Fung}},
  \bibinfo{author}{\bibfnamefont{A.}~\bibnamefont{Kumar}}, \bibnamefont{and}
  \bibinfo{author}{\bibfnamefont{N.~X.} \bibnamefont{Fang}},
\bibinfo {title} {Electron-photon scattering mediated by localized plasmons: A quantitative analysis by eigen-response theory},
  \bibinfo{journal}{Phys. Rev. B} \textbf{\bibinfo{volume}{89}},
  \bibinfo{pages}{045408} (\bibinfo{year}{2014}).

\bibitem[{\citenamefont{Bohren and Huffman}(2008)}]{Bohren}
\bibinfo{author}{\bibfnamefont{C.~F.} \bibnamefont{Bohren}} \bibnamefont{and}
  \bibinfo{author}{\bibfnamefont{D.~R.} \bibnamefont{Huffman}},
  \emph{\bibinfo{title}{Absorption and scattering of light by small particles}}
  (\bibinfo{publisher}{John Wiley \& Sons}, \bibinfo{year}{2008}).

\bibitem[{\citenamefont{Wang et~al.}(2020)\citenamefont{Wang, Lee, Dong, Dong,
  Yu, and Fung}}]{Jin2}
\bibinfo{author}{\bibfnamefont{J.}~\bibnamefont{Wang}},
  \bibinfo{author}{\bibfnamefont{K.~F.} \bibnamefont{Lee}},
  \bibinfo{author}{\bibfnamefont{H.~Y.} \bibnamefont{Dong}},
  \bibinfo{author}{\bibfnamefont{Z.-G.} \bibnamefont{Dong}},
  \bibinfo{author}{\bibfnamefont{S.~F.} \bibnamefont{Yu}}, \bibnamefont{and}
  \bibinfo{author}{\bibfnamefont{K.~H.} \bibnamefont{Fung}},
\bibinfo {title} {Collective resonances in a circular array of gyromagnetic rods},
  \bibinfo{journal}{Phys. Rev. B} \textbf{\bibinfo{volume}{101}},
  \bibinfo{pages}{045425} (\bibinfo{year}{2020}).

\bibitem[{\citenamefont{Yasumoto}(2005)}]{Yasu}
\bibinfo{author}{\bibfnamefont{K.}~\bibnamefont{Yasumoto}},
  \emph{\bibinfo{title}{Electromagnetic theory and applications for photonic
  crystals}} (\bibinfo{publisher}{CRC press}, \bibinfo{year}{2005}).

\bibitem[{\citenamefont{Botten et~al.}(2000)\citenamefont{Botten, Nicorovici,
  Asatryan, McPhedran, de~Sterke, and Robinson}}]{Botten}
\bibinfo{author}{\bibfnamefont{L.~C.} \bibnamefont{Botten}},
  \bibinfo{author}{\bibfnamefont{N.-A.~P.} \bibnamefont{Nicorovici}},
  \bibinfo{author}{\bibfnamefont{A.~A.} \bibnamefont{Asatryan}},
  \bibinfo{author}{\bibfnamefont{R.~C.} \bibnamefont{McPhedran}},
  \bibinfo{author}{\bibfnamefont{C.~M.} \bibnamefont{de~Sterke}},
  \bibnamefont{and} \bibinfo{author}{\bibfnamefont{P.~A.}
  \bibnamefont{Robinson}},
\bibinfo {title} {Formulation for electromagnetic scattering and propagation through grating stacks of metallic and dielectric cylinders for photonic crystal calculations. part i. method},
 \bibinfo{journal}{J. Opt. Soc. Am. A}
  \textbf{\bibinfo{volume}{17}}, \bibinfo{pages}{2165} (\bibinfo{year}{2000}).

\bibitem[{Sup({\natexlab{b}})}]{Supp0}
\bibinfo{note}{See Fig. S1(a) of Supplemental Material at [URL will be inserted
  by publisher]}.

\bibitem[{\citenamefont{Dong et~al.}(2013{\natexlab{a}})\citenamefont{Dong,
  Wang, and Cui}}]{Dong}
\bibinfo{author}{\bibfnamefont{H.~Y.} \bibnamefont{Dong}},
  \bibinfo{author}{\bibfnamefont{J.}~\bibnamefont{Wang}}, \bibnamefont{and}
  \bibinfo{author}{\bibfnamefont{T.~J.} \bibnamefont{Cui}},
\bibinfo {title} {One-way tamm plasmon polaritons at the interface between magnetophotonic crystals and conducting metal oxides},
  \bibinfo{journal}{Phys. Rev. B} \textbf{\bibinfo{volume}{87}},
  \bibinfo{pages}{045406} (\bibinfo{year}{2013}{\natexlab{a}}).

\bibitem[{\citenamefont{Dong et~al.}(2013{\natexlab{b}})\citenamefont{Dong,
  Wang, and Fung}}]{Dong2}
\bibinfo{author}{\bibfnamefont{H.~Y.} \bibnamefont{Dong}},
  \bibinfo{author}{\bibfnamefont{J.}~\bibnamefont{Wang}}, \bibnamefont{and}
  \bibinfo{author}{\bibfnamefont{K.~H.} \bibnamefont{Fung}},
\bibinfo {title} {One-way optical tunneling induced by nonreciprocal dispersion of tamm states in magnetophotonic crystals},
  \bibinfo{journal}{Opt. Lett.} \textbf{\bibinfo{volume}{38}},
  \bibinfo{pages}{5232} (\bibinfo{year}{2013}{\natexlab{b}}).

\bibitem[{\citenamefont{Bulgakov and Sadreev}(2014)}]{PRA}
\bibinfo{author}{\bibfnamefont{E.~N.} \bibnamefont{Bulgakov}} \bibnamefont{and}
  \bibinfo{author}{\bibfnamefont{A.~F.} \bibnamefont{Sadreev}},
\bibinfo {title}{Bloch bound states in the radiation continuum in a periodic array of dielectric rods},
  \bibinfo{journal}{Phys. Rev. A} \textbf{\bibinfo{volume}{90}},
  \bibinfo{pages}{053801} (\bibinfo{year}{2014}).

\end{thebibliography}
\end{document}